# High optical magnetism of dodecahedral plasmonic meta-atoms.

V. Many[1], R. Dezert[2],  E. Duguet[1], A. Baron[2], V. Jangid[2], V. Ponsinet[2], S. Ravaine[2], P. Richetti[2], P. Barois[2], M. Tréguer-Delapierre[1]

[1]ICMCB, CNRS & Univ. Bordeaux

[2]CRPP & Univ. Bordeaux

**Abstract:** The generation in artificial composites of a magnetic response to light comparable in magnitude with the natural electric response, may offer an invaluable control parameter for a fine steering of light at the nanoscale. In many experimental realizations however, the magnetic response of artificial meta-atoms is too weak so that there is a need for new designs with increased magnetic polarizability. Numerical simulations show that geometrical plasmonic nanostructures based on the ideal model of Platonic solids are excellent candidates for the production of strong optical magnetism in visible light. Inspired by this model, we developed a bottom-up approach to synthesize plasmonic nano-clusters made of twelve gold patched located at the center of the faces of a dodecahedron. The scattering of the electric and magnetic dipole induced by light are measured across the whole visible range. The ratio of the magnetic to electric response at resonance is found three times higher than its counterpart measured on disordered plasmonic clusters ("plasmonic raspberries") of the same size. Numerical simulations confirm the experimental measurements of the magnetic response.

**Introduction:**

The search for artificial optical magnetism in engineered composites has attracted intense research efforts since the emergence of the concept of metamaterials in 2000 [1]. While the optical properties of conventional materials or systems are governed by the sole electric polarizability, controlling the magnetic polarizability would offer an additional lever, potentially enabling a full control of the amplitude and the phase of a light wave transmitted through and reflected from an interface. Many applications of bulk materials or optical films made of magneto-electric elements have been proposed by theory such as total transmission or reflection (Huyghens' surfaces), band-pass filtering, pulse compression, directional steering of light, polarization beam splitters, converters and analyzers to name a few [2]. In natural materials, the vanishing of the induction of a magnetization by an external magnetic field at optical frequencies is well known [3]. Nevertheless, several types of artificial nanostructures have been proposed to bypass this rule. They are generally based on the design of optically resonant building blocks (meta-atoms) of sub-wavelength size: the resonance warrants an efficient optical response whereas the sub-wavelength size of the meta-atoms is required for a homogeneous electromagnetic response of the metamaterial or metasurface made out of them. The proposed nanostructures fall essentially into two classes, namely plasmonic nanoclusters and high-index Mie resonators. In the first family, a set of metallic nano-objects is organized in pairs [4] or larger clusters [5-8] so as to enable the excitation of loops of plasmonic currents which generate a magnetization that oscillates at the frequency of the impinging light. In the second family, the magnetic response to light is obtained at the magnetic Mie resonances of meta-atoms of simple shapes like spheres or cylinders [9]. The resonance condition is reached when the wavelength inside the meta-atom matches its size, which, together with the requirement of sub-wavelength size imposes a high refractive material. Crystalline silicon is an excellent candidate for its large refractive index (~4) with low losses [9]. For both families, many experimental realizations of magnetic devices were produced by powerful "top-down" techniques such as lithography (including UV, electron-beam and nano-imprint lithography), laser printing or ion etching. The materials were mostly noble metals (Au or Ag) but also transparent conducting oxide or doped semiconductors in the first family [10], non-doped semiconductors (silicon, germanium, tellurium) in the second family [9]. The top-down approach provides unequalled spatial resolution for surface organization of nano-sized elements. Its applications to silicon devices are particularly attractive as they present low losses, they are CMOS compatible and the frequency of the electric and magnetic resonances can be tuned independently by a proper design of the shape of the meta-atoms [11]. Although the top-down techniques are not limited to small areas or purely 2D systems [12], their implementation becomes increasingly difficult as the size and the thickness of the optical devices increase. An alternative "bottom-up" approach was proposed which combines nano-chemistry for the large-scale production of magnetic meta-atoms and self-assembly for the fabrication of thick materials [13,14] or thin films of large area [15]. The model of isotropic magnetic nanoclusters or "plasmonic raspberries" proposed by Simovski and Tretyakov [8] in particular has stimulated a lot of synthetic efforts. Nanoclusters exhibiting a strong magnetic response were indeed successfully synthesized by several authors [16-21] and a significant variation of the effective magnetic permeability was demonstrated in a self-assembled bulk magnetic materials [20]. While being quite valuable on a fundamental point of view, these experimental works have evidenced the need to enhance the magnetic response in order to reach a usable optical functionality. The highest magnetic to electric ratio of the dipolar response measured on plasmonic raspberries with silver satellites is of order 0.3 [20] a record value for meta-atoms so far, but still far from the ratio of 1 required for reflective Huyghens' metasurfaces [11,22]. Variations of the real part of the relative magnetic permeability of the bulk magnetic material assembled in ref. [20] span an interval of 0.80 to 1.45 whereas possible applications to super-lensing, wave-front shaping or impedance matching require negative, near zero or high values of the permeability [23,24]. Once again, nanoresonators made out of crystalline silicon prove to be highly efficient [25]. Unfortunately, silicon nanospheres of well-controlled size and crystallinity are not easily available in macroscopic amounts yet. A fair number of silicon nanospheres can be prepared and crystallized by intense laser pulses [26] but for larger amounts, a productive synthetic route has still to be found [27].

We focus in this paper on the improvement of the magnetic response of plasmonic nanoclusters which are accessible to the methods of nano-chemistry. However, instead of improving the structural parameters of the plasmonic raspberry model based on a large number of metallic satellites randomly distributed around a dielectric core [21], we investigate an alternative model based on a determined number of metallic satellites precisely located at the center of the faces of a Platonic solid, i.e. a convex polyhedron with 4, 6, 8, 12 or 20 faces. An example with six satellites sitting on the faces of a cube was given in reference [5]. We choose here the so-called dodecapod structure based on the dodecahedron Platonic solid with twelve faces. The main results of the paper are (i) that gold dodecapods (AuDDPs) can be synthesized in large amount with a high purity and (ii) that the magnetic to electric polarization ratio measured on these new meta-atoms is three times higher than the reference value measured on plasmonic raspberries of similar size [18].

The paper is organized as follows: the numerical simulations that guided this work are presented in sections I. Section II is devoted to the synthesis of the dodecapods, which is based on the concept of "patchy particles" [28]. The optical study is presented in section III, followed in IV by numerical simulations of the synthesized objects and a conclusion.

**I – Numerical simulations of plasmonic platonic solids:**

Several experimental realizations [16-21] have demonstrated the efficiency of the Simovski-Tretyakov model of plasmonic raspberries [8] for the generation of a magnetic polarization in visible light. In this model, the metallic satellites are randomly distributed around the core: neither their exact number, not their precise position are controlled. On the other hand, the electromagnetic coupling between the plasmonic satellites, and hence the magnitude of the current ring generating the magnetic dipole, depends critically on the position of the satellites relative to one another. One may then wonder whether a stronger magnetic mode could be reached with a finer control of the position of the satellites. This question might have remained virtual until Chomette et al. realized the synthesis of multipod plasmonic clusters based on the geometrical model of Platonic solids decorated with four (tetrapods), six (hexapods) and twelve (dodecapods) satellites [28]. In this section, we investigate numerically the scattering properties of perfect multipods in order to define the most efficient morphology and provide guidance for nanochemists. The four studied morphologies are shown in figure 1abcd.

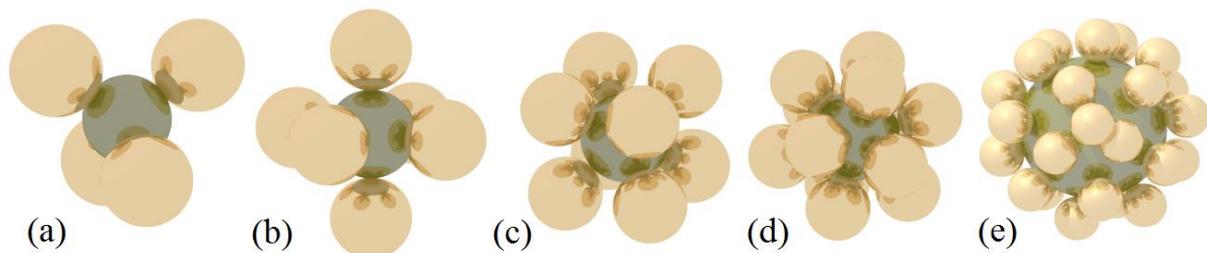

*Figure 1: Sketch of the tetrapod (a), hexapod (b), octapod (c) and dodecapod (d) models used in numerical simulations with the same amount of gold distributed in N satellites around the same dielectric core. The radius of the satellites scales as $N^{-1/3}$ (e) The "plasmonic raspberry" model [8,18].*

In order to identify the sole influence of the morphology, the scattering properties of the four multipods (referred to as N-pods with N = 4, 6, 8 and 12) are computed keeping constant the radius of the core (50 nm) and the total volume of gold, which implies that the radius of the N satellites scales as $N^{-1/3}$. We employed a T-Matrix code developed by Mackowski [29] providing the solution to Maxwell's equations for an ensemble of non-overlapping spheres. Following the formalism of the Mie theory describing the scattering of a single sphere [30], the elements of the scattering matrix of the N-pods are expanded in series of scattering coefficients $a_n^{N-pod}$ and $b_n^{N-pod}$. The first order coefficients $a_1^{N-pod}$ and $b_1^{N-pod}$ correspond respectively to the electric (ED) and magnetic (MD) dipolar mode, $a_2^{N-pod}$ and $b_2^{N-pod}$ to the electric (EQ) and magnetic (MQ) quadrupolar modes and so on. Figure 2a shows the scattering efficiency of the different modes computed for a dodecapod (N=12).

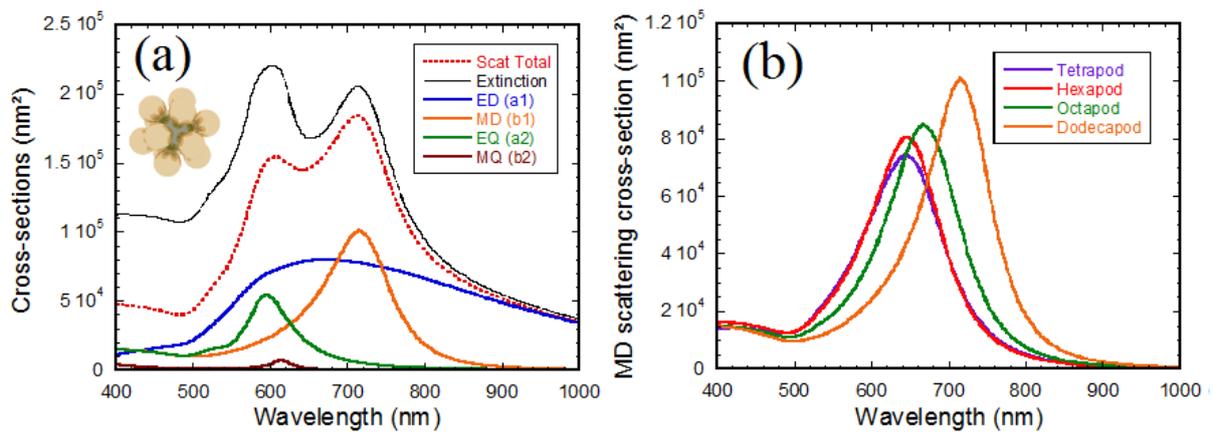

*Figure 2:* (a) Extinction and scattering cross-sections of a dodecapod in water. The radii of the silica core and of the gold satellites are 50 and 40 nm respectively. The scattering is essentially dominated by the three contributions of the electric dipole (ED), magnetic dipole (MD) and electric quadrupole (EQ). (b) Magnetic (MD) scattering cross-section computed for the four N-pods of figure 1a-d. The radii R(N) of the satellites are 57.7, 50.4, 45.8 and 40 nm for N=4,6,8,12 respectively to ensure a constant volume of gold $N.4/3 \pi R(N)^3$.

The scattering results essentially from three contributions namely ED, MD and EQ. The MQ mode is much weaker and higher order modes (not shown in figure 2a) are negligible. The MD and ED scattering are comparable in strength in a broad band around 720 nm. Figure 2b shows the sole MD contribution of the four multipods of figure 1. The dodecapod structure yields the highest MD scattering. The MD scattering efficiency, defined as the scattering cross-section normalized by the geometrical cross-section, reaches 1.88. The dodecapod structure is therefore chosen as the most desirable for the optimization of the magnetic response.

**II – Synthesis:**

Guided by the numerical simulations reported in section I, we implemented a synthetic method targeting the ideal dodecapod structure of Fig. 1d. Isotropic 3D plasmonic clusters were prepared by a multistep synthetic process from silica/polystyrene dodecahedral templates (Fig. S1) as summarized in Figure 3. Details of the synthesis have recently been published [28]. The last three steps of the synthetic pathway shown in Fig.3 are the stages of regioselective gold seeds adsorption, seed-mediated growth on the dielectric template and condensation of the gold domains. Figure 4 shows typical dark field micrographs of the nano-objects obtained via this approach. The electron micrographs as well as the elemental mapping evidence the well-defined patterns of gold, symmetrically arranged around the silica. After the seed-mediated growth step, the twelve gold deposits consist of agglomerates of nanoparticles (Fig.4a-b). Annealing at 220°C for a couple of minutes in a polyol transforms the agglomerates into smooth, dense particles (Fig.4c-d). The size of the gold particles can be adjusted by increasing the amount of gold precursor during the seed-mediated growth stage. A strong advantage of this approach is that the monodisperse golden dodecapods are produced, in high yield, at the gram scale.

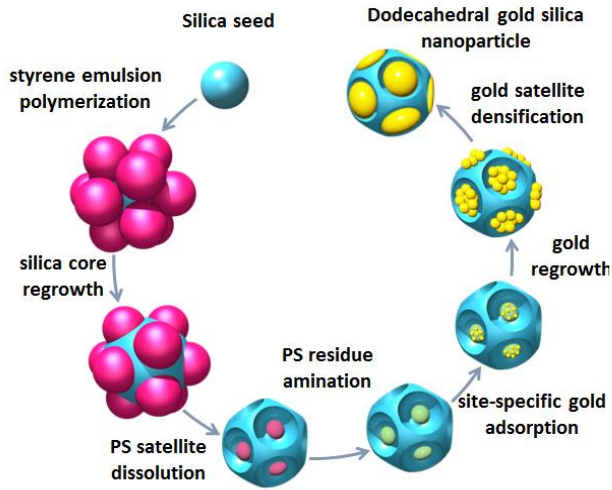

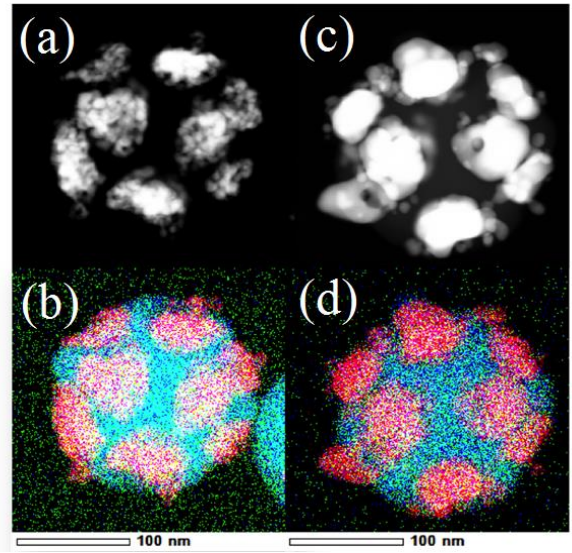

*Figure 3:* Schematic process for the synthesis of gold dodecapods. The growth of the polystyrene nodules defines twelwe geometrical sites which are subsequently functionalized to promote gold adsorption. The final structure is obtained after gold regrowth and thermal annealing.

*Figure 4:* (a,c) Scanning Electron micrographs of dodecapods S4 before (a) and after(c) condensation. (c,d) Elemental analysis revealing silica (blue) and gold (red). Images (a,b) and (b,c) correspond to samples S3-NC and S3-C respectively (see text).

Three dodecapod samples named S1 to S3 with increasing amount of gold were synthesized (cf Supporting information section for experimental details). Optical studies were performed before and after condensation of the satellites by thermal annealing. Figure S4 shows extinction measurements performed on the two sets of four samples. Suffixes NC and C stand for non-condensed and condensed, respectively before and after annealing. Extinctions curves show broad maxima, consistent with plasmonic resonances. The maxima red-shift and broaden upon increasing the amount of gold.

### III – Optical studies: **Static Light Scattering (SLS):**

Following the experimental method first implemented for the study of gold raspberries [18], the scattering properties of the gold dodecapods were studied by polarization resolved static light scattering sketched in Fig. S5. The scattering angle $\theta$ is set at 90°. The incident field $\mathbf{E}_i$ propagating along z is linearly polarized in the (x,y) plane at angle ϕ from the x axis perpendicular to the scattering plane (y,z).

We use the amplitude scattering matrix formalism for the analysis of the scattering data [30]:

$$\begin{pmatrix} E_{//S} \\ E_{\perp S} \end{pmatrix} = \frac{e^{ikr}}{-ikr} \begin{pmatrix} S_2 & S_4 \\ S_3 & S_1 \end{pmatrix} \begin{pmatrix} E_i \cos\varphi \\ E_i \sin\varphi \end{pmatrix} \quad (1)$$

$E_{//S}$ and $E_{\perp S}$ are the components of the scattered field parallel and perpendicular to the scattering plane. In the case of an isotropic scatterer, which we assume here, the $S_i$ coefficients do not depend on $\varphi$ and the non-diagonal terms $S_3$ and $S_4$ vanish. The scattered field can be formally expanded over vector spherical harmonics of increasing order n leading to the following expansions of $S_1$ and $S_2$:

$$S_1 = \sum_n \frac{2n+1}{n(n+1)}(a_n \pi_n + b_n \tau_n)$$
$$S_2 = \sum_n \frac{2n+1}{n(n+1)}(a_n \tau_n + b_n \pi_n)$$
(2)

$a_n$ and $b_n$ reduce to the standard Mie scattering coefficients for spheres. For dodecapods, they are computed via the T-Matrix expansion described in section II. The angle-dependent coefficients $\pi_n$ and $\tau_n$ constructed from the Legendre polynomials are alternately odd and even functions of $\cos\theta$ so that for $\theta = 90°$, each scattering coefficient $a_n$ and $b_n$ appears either in $S_1$ or $S_2$ but never in both. As shown in section II, the scattering amplitude $S_1$ reduces to the radiation of the electric dipole ED ($a_1$) only whereas $S_2$ cumulates the radiations of the magnetic dipole MD ($b_1$) and the electric quadrupole EQ ($a_2$) [31, 18,20].

The intensities measured along the two output polarizations for isotropic scatterers then read:

$$I_{\perp S} = I_0(\lambda) N_{DDP} \frac{|S_1(\theta=90°)|^2}{k^2} \delta\Omega g(\lambda, \delta\Omega) \cos^2\varphi = A_\perp(\lambda)\cos^2\varphi \quad (3a)$$

$$I_{//S} = I_0(\lambda) N_{DDP} \frac{|S_2(\theta=90°)|^2}{k^2} \delta\Omega g(\lambda, \delta\Omega) \sin^2\varphi = A_{//}(\lambda)\sin^2\varphi \quad (3b)$$

in which $I_0(\lambda)$ is the spectral irradiance of the incident beam, $N_{DDP}$ is the number of gold dodecapods in the scattering volume, and $\delta\Omega$ is the solid angle of the detection window. $g(\lambda, \delta\Omega)$ is an unknown function that accounts for the spectral sensitivity of the detector and optical transmission or reflection of all optical elements.

Experimentally, a supercontinuum laser white source (SuperK EXB-6 with SuperK Split UV-visible filter from NKT Photonics) was used to deliver a white light beam in the range 400-900 nm on to a dilute aqueous suspension of dodecapods. The linear polarization of the incident light was set by a Glan-Taylor polarizer (Thorlabs GL5-A). A Fresnel Rhomb achromatic half-wave retarder (Thorlabs FR600HM) mounted on a motorized rotation stage was used to rotate the incident polarization by an angle $\varphi$. The scattered light was collected at a fixed scattering angle $\theta = 90°$ for the two output polarizations perpendicular (signal $I_{\perp S}$) and parallel ($I_{//S}$) to the scattering plane and delivered through a collimated fiber (NA = 0.25) to a mini-spectrometer (Hamamatsu model C10083CA) for spectral analysis.

Figure 5 illustrates the data collected on sample S3-C. The perpendicular and parallel signals $I_{\perp S}$ and $I_{//S}$ plotted at a wavelength of 640.5 nm, are fitted to simple functions $A_\perp \cos^2\varphi + B$ and $A_{//}\sin^2\varphi + B$ respectively. The background signal B, identical for the two functions, does not exceed a few percent (2 to 7.5% across the wavelength range) of the amplitude $A_\perp$, which supports the assumption of isotropic scatterer [32].

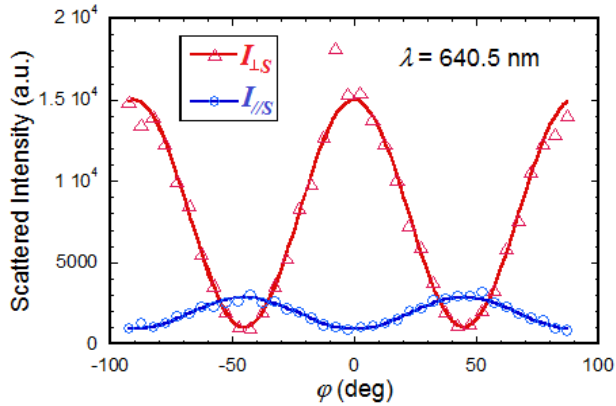

**Figure 5:** *Example shown at λ=640.5 nm of the signal scattered by the sample S3-C along the two output polarization channels $I_{\perp S}$ and $I_{//S}$ perpendicular and parallel to the scattering plane respectively. Solid lines are a fit to cos²φ and sin²φ functions.*

Taking the ratio of the perpendicular to parallel signal $A_\perp / A_{//}$ (eqs.3) eliminates all the unknown functions in eqs. 3a,b and provides an estimate of the strength of the magnetic dipole MD relative to ED:

$$A_\perp / A_{//} = \frac{|S_2(\theta=90°)|^2}{|S_1(\theta=90°)|^2} = \frac{|3b_1 - 5a_2|^2}{|3a_1 - 5b_2|^2} \equiv \frac{MD + EQ}{ED + MQ} \approx \frac{MD}{ED} \quad (4)$$

This indicator of the MD/ED ratio is plotted vs. wavelength for all dodecapod samples in figure 6. It exhibits for all samples a broad maximum in the visible range. The maximum increases and red-shifts upon increasing the amount of gold from ~5% to ~9% for non-condensed dodecapods. Condensation of the satellites at constant amount of gold further increases the maxima from ~9% to ~14%. The data measured on gold raspberries of similar size [18] are also shown in Fig.6 for comparison. Figure 6 reveals a spectacular improvement of the ratio from 4.2% for gold raspberries to 14% for dodecapods.

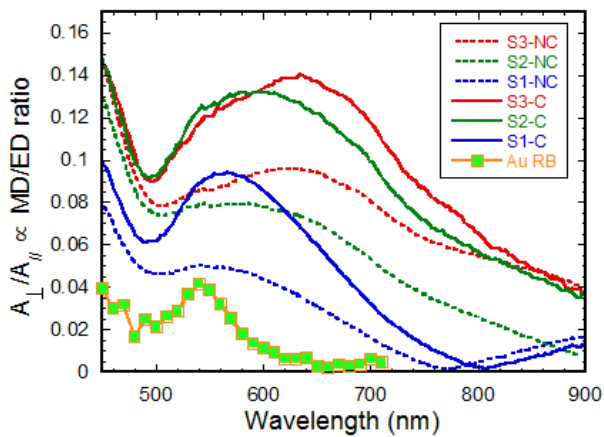

**Figure 6**: *Plots of the experimental function $A_\perp/A_{//}$ as an estimator of the ratio of the magnetic to electric dipolar scattering. Dashed and solid lines correspond to light scattering measurements respectively before and after condensation. The ratio increases with gold load and with condensation, well above values observed on gold raspberries shown for comparison (squares, from ref. [18]).*

The parallel MD+EQ and perpendicular ED scattering cross-sections along the two polarization channels can be extracted separately by normalizing the data with a reference dispersion of calibrated silica nanoparticles ($D_{SiO2}$ = 95 nm) in water [18]. The scattered intensity from the reference sample is similar to Eq. (3a) by replacing $|S_1(\theta = 90°)|^2/k^2$ with the differential scattering cross section $\sigma_{ref}^{\theta=\pi/2,\varphi=0}$ of a silica particle in water at scattering angles $\theta = \pi/2$, $\varphi = 0$, which is easily calculated from the Mie theory and scales as $1/\lambda^4$. The dodecapods and silica signals were collected in the same experimental conditions so that the quantities $I_0$, $\delta\Omega$, and $g(\lambda, \delta\Omega)$ are the same for dodecapods and for the reference. Dividing eq. (3b) for dodecapods by eq. (3a) written for the silica reference yields the differential scattering cross sections of dodecapods along the particular directions $\theta = \varphi = 90°$:

$$\sigma_{//}^{MD+EQ}(\theta=\varphi=\pi/2) = \frac{|S_2(\theta=\pi/2)|^2}{k^2} = K \frac{A_{//}(\lambda)}{A_\perp^{ref}(\lambda)} \sigma_{ref}^{\theta=\pi/2,\varphi=0}(\lambda) \quad (5)$$

Superscript MD+EQ reminds that $\sigma_{//}$ is dominated by the added scattering contribution of the magnetic dipole and electric quadrupole. Figure 7a shows plots of the parallel scattering cross-sections $\sigma_{//}^{MD+EQ}(\theta=\varphi=\pi/2)$ for three non-condensed dodecapods of increasing gold load. All three curves exhibit a clear resonance which red-shifts and broadens with increasing gold amount. Due to a large uncertainty in the concentration of the dodecapods after synthesis and purification, the concentration ratio $K = N_{ref}/N_{DDP}$, independent of wavelength, could not be determined accurately for each sample. To avoid any misleading interpretation, the resonance maxima are normalized to 1. Annealing sharpens the resonances with a slight red-shift in wavelength (Fig.7b).

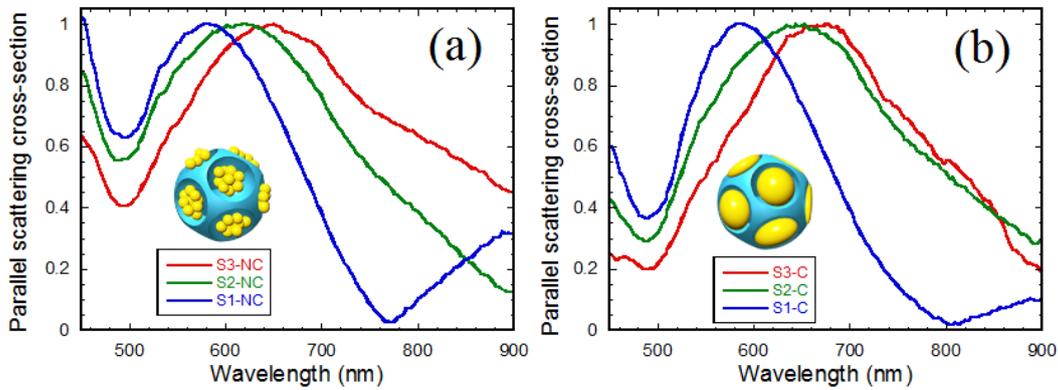

**Figure 7:** *Parallel scattering cross-section $\sigma_{//}^{MD+EQ}$ (eq.5, maximum normalized to 1) of the dodecapod samples before annealing (a) and after (b). The resonant response red-shifts with increasing gold load. Annealing sharpens the resonance, with very slight red-shift in wavelength.*

### IV – Numerical simulations:

The condensed dodecapods have proven to generate a MD/ED ratio about three times higher than gold raspberries of comparable size. This is a clear success of the new morphology. Nevertheless, the synthesized dodecapods are not as perfect as the ideal guiding model of Figure 1d. We refined our numerical simulations in order to capture the main features of the optical behavior summarized in figures 6 and 7 and to try and understand the morphological features that cause the differences in behavior compared to the ideal case. Two different structural models are used to compute the scattering properties of the dodecapods before (non-condensed) and after (condensed) annealing.

**IV-a: Non condensed dodecapods**: As shown in Fig.4, the satellites before annealing consist of agglomerates of gold nanoparticles. We mimic the non-condensed dodecapods as shown in the inset of figure 10a. Twelve satellite sphres of radius $R_{sat}$ are partially filled with spherical gold nanoparticles of radius $R_{Au}$ inside the volume of a truncation sphere of radius $R_{trunc}$. Increasing the amount of gold from sample S1-NC to S3-NC may be produced by increasing $R_{trunc}$ (bigger agglomerate) and/or increasing $R_{Au}$ while keeping the number of gold NPs constant (higher volume fraction of gold within each agglomerate).

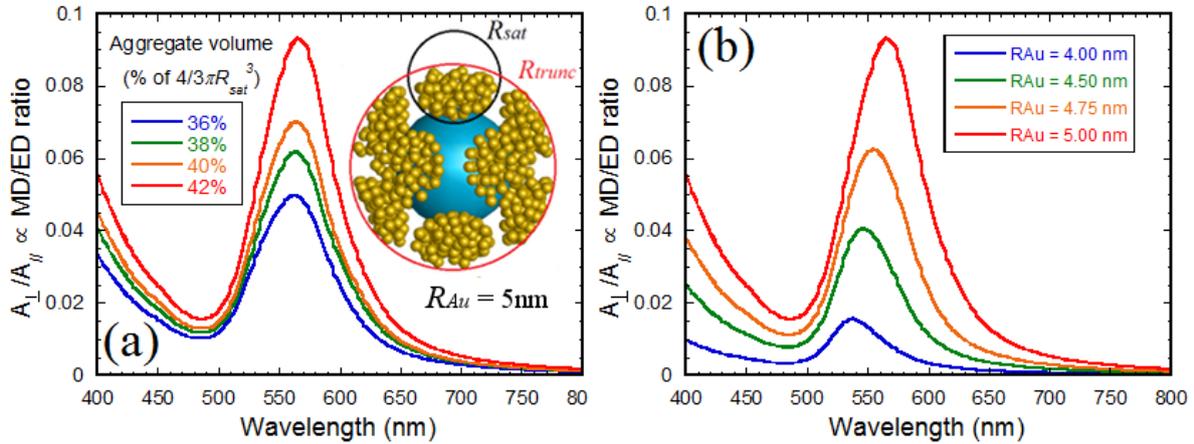

*Figure 8: (a) Plot of the computed magnetic to electric estimator $A_\perp/A_\parallel$ vs. wavelength for different volumes of gold aggregates in non-condensed dodecapods. The cartoon illustrates the construction of the structure: the volume of the aggregates is defined as the intersection of twelve spherical satellites of radius $R_{sat}$ (black circle) with a truncation sphere of radius $R_{trunc}$ (red circle). These volumes are filled with non-intersecting gold nanospheres of radius $R_{Au} = 5$ nm with a volume fraction set to 40%. The volume of the aggregates, controlled by $R_{trunc}$, is given as a percentage of the volume of the full satellite sphere $4/3\pi R_{sat}^3$. (b) Same $A_\perp/A_\parallel$ ratio computed at constant aggregate size, but for different radii $R_{Au}$. Note the red-shift of the magnetic resonances with increasing size of the gold nanoparticles.*

Figures 10a,b show that the experimental values of the magnetic to electric ratio $A_\perp/A_\parallel$ (Fig.8, dashed lines) are very well reproduced by the model of aggregates of gold nanoparticles of radius 5 nm. Increasing the amount of gold in the model either by increasing the volume of the aggregate (via $R_{trunc}$) or the size of the gold nanoparticles ($R_{Au}$) enhances the magnetic response as observed experimentally. However, the experimental red-shift observed upon increasing the amount of gold is only reproduced by increasing the size of the gold nanoparticles within the aggregates, which in turn reduces their average separation. This effect is consistent with the well-known plasmon ruler law stating that the resonance wavelength of a set of gold nanoparticles red-shifts as the separation of the nanoparticles is reduced [33]. This observation indicates that the size, and not just the number, of the gold grains increases upon successive regrowth steps.

**IV-b: Condensed dodecapods**:
Four systems are considered, one for each volume of gold considered in the non-condensed case shown in Fig. 8(a). The patches are obtained by keeping the intersections of a core-sphere of radius $R'_{trunc}$ with satellite spheres of radius $R_{sat}$. Of course $R'_{trunc}$ is not equal to $R_{trunc}$ as it set so as to conserve the volume of gold in the system. An illustration of a typical system considered is shown in inset on Fig. 9(a). The scattering properties of the system are calculated using the finite element method based commercial software COMSOL Multiphysics in the scattering formalism. The field scattered by the patchy dodecapod is computed and decomposed on a spherical harmonic basis using the method of volume currents [34]. As a result, the spectral dependence of the $a_i$ and $b_i$ coefficients are determined, which in turn enables the calculation of the scattering coefficients $S_1$ and $S_2$. Figure 9(a) shows how $A_\perp/A_\parallel$ evolves as a function of wavelength. We see a clear increase in the amplitude of this magnetic to electric ratio as well as a red-shift in the resonance wavelength compared to the non-condensed case (see Fig. 8(a)), in accordance with the experimental observations.

The experimental behavior is qualitatively reproduced and we may discuss the reasons for which differences still remain, namely that the maximum value of $A_\perp/A_\parallel$ (~0.10-0.14) is significantly lower compared to the numerical case (~1) and the red-shift is not as large. The annealing process actually dewets the gold such that the shape does not necessarily conform to the same shape – intersection of two spheres resulting in a lens-like shape – as that which we simulate. This means that each patch will take on a shape that is different to the lens shape and probably more spheroidal. In the end the simulated patches are thinner than the experimental ones resulting in larger aspect ratios which are known to resonate at higher wavelengths compared to smaller aspect ratios [35]. Furthermore, the lens-shape is actually sharp along the edges, which will result in larger local fields than in the experimental case. Consequently the capacitive effect is increased between patches, which will produce larger polarization currents and result in a larger magnetic dipole moment, hence the higher values of the magnetic to electric scattering ratio.

Getting a more refined model is certainly meaningful in the future to better guide the nano-synthesis but is beyond the scope of this paper and we find that our current understanding of the system is sufficient for a gross road-map for improvement. In essence the numerical system actually has great potential as its scattering properties are very good. If the nano-synthesis could be made to preserve the aspect ratio, the patchy-dodecapod system would make an excellent magneto-optical resonator. As a matter of fact, as shown by Fig. 9(b), condensation drastically increases the amplitude of the scattering-cross sections of all multipoles present, but the magnetic dipole is the resonance for which the effect is most dramatic. This reveals the positive role of condensation of the magneto-optical properties of such systems.

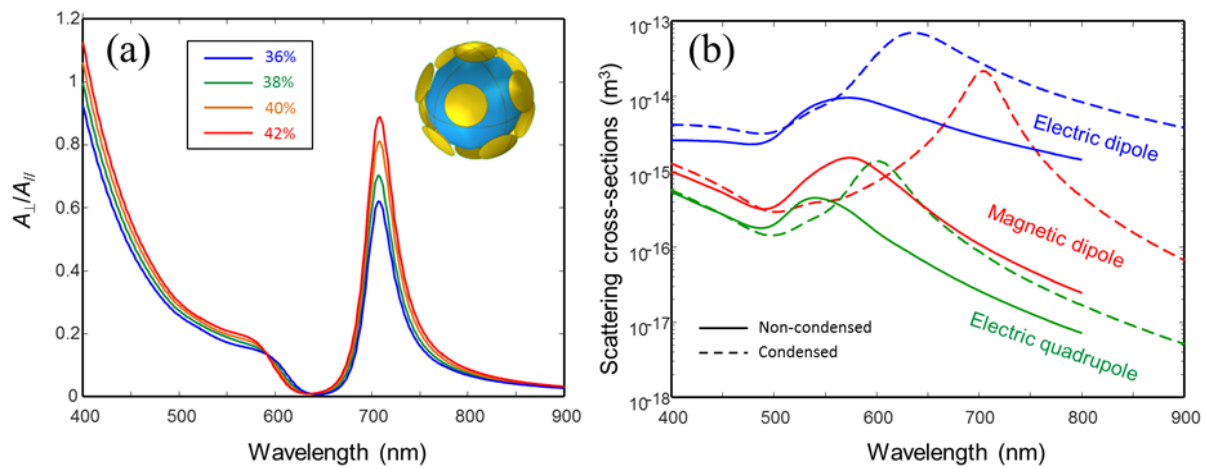

*Figure 9:* (a) Calculated plots of the computed magnetic to electric estimator $A_\perp/A_{//}$ vs. wavelength for different volumes of gold aggregates in condensed dodecapods. The inset in the top-right corner represents a typical model of the patchy dodecapod used in the finite element simulation. The percentages in the legend are those taken from Fig. 10(a) and the objects have identical volumes of gold and identical values of the core particles as well as $R_{Au}$. (b) Simulated scattering cross-sections of the electric (blue) and magnetic dipoles (red) as well as the electric quadrupole (green) in both the non-condensed (continuous line) and condensed (dashed line) cases.

## VI – Conclusion:

The main result of this work is the demonstration that the experimental ratio of the magnetic to electric response of the gold-dodecapods (MD/ED=14%) is more than three time higher than its counterpart measured in gold raspberries of similar size (4.5%). As argued in the introduction, the magnetic response to light has an impact on optical functionality if its magnitude is similar to the electric response, i.e. if the ratio is about 1. Gold dodecapods are still below this number. It has been shown however that changing gold for silver reduces the optical losses and strongly enhances the MD/ED ratio of plasmonic raspberries. Up to now, the highest MD/ED ratio obtained with silver raspberries is about 0.28 [20]. Gaining a factor 3 over silver raspberries is then fully relevant to reach the target MD/ED~1. Future work will definitely have to explore the synthesis of silver-dodecapods.

In conclusion, we claim that the present work validates the superiority of the dodecapod structure over the disordered distribution of a larger number of smaller satellites in the raspberry model. The well-defined positions of the satellites indeed offer a better control of the collective electromagnetic coupling which determines the strength of the circular plasmonic currents generating the magnetization.

**Supplementary Material:** Experimental details, TEM images of particles during the different stages of fabrication, extinction spectra of the dodecahedral plasmonic particles before and after densification of the satellites.

**Acknowledgments**
TEM, HR-TEM and HR-SEM experiments were performed at the Plateforme de Caractérisation des Matériaux of the University of Bordeaux and the Plateforme Castaing of the University of Toulouse. This work was supported by the LabEx AMADEus (ANR-10-

**References:**

[1] Smith D.R., Padilla W.J., Vier D.C. et al. Composite Medium with Simultaneously Negative Permeability and Permittivity. Phys. Rev. Lett., 2000, 84, 4184-4187.

[2] Glybovski S.B., Tretyakov S.A., Belov P.A., Kivshara Y.S. and Simovski C.R. Metasurfaces: From microwaves to visible. Physics Reports 2016, 634, 1–72.

[3] Landau L.D. and Lifshitz E.M., Electrodynamics of Continuous Media, Pergamon Press, Oxford, 1960.

[4] Shalaev V., Optical negative-index metamaterials. Nature Photonics 2007, 1, 41–48.

[5] Alù A. and Engheta N., The quest for magnetic plasmons at optical frequencies. Opt. Express 2009, 17, 5723-5730.

[6] Alù A., Salandrino A. and Engheta N., Negative effective permeability and left-handed materials at optical frequencies. Opt. Express 2006, 14, 1557-1567.

[7] Alù A. and Engheta N. Dynamical theory of artificial optical magnetism produced by rings of plasmonic nanoparticles. Phys. Rev. B 2008, 78, 085112.

[8] Simovski C.R. and Tretyakov S.A. Model of isotropic resonant magnetism in the visible range based on core-shell clusters. Phys. Rev. B, 2009, 79, 045111.

[9] Jahani S. and Jacob Z. All-dielectric metamaterials. Nature Nanotech. 2016, 11, 23-36.

[10] Naik G.V., Shalaev V.M. and Boltasseva A. Alternative Plasmonic Materials: Beyond Gold and Silver. Adv. Mater. 2013, 25, 3264–3294.

[11] Staude I., Miroshnichenko A.E., Decker M., et al. Tailoring Directional Scattering through Magnetic and Electric Resonances in Subwavelength Silicon Nanodisks. ACS Nano 2013, 7, 7824-7832.

[12] Walia S., Shah C.M., Gutruf P. et al. Flexible metasurfaces and metamaterials: A review of materials and fabrication processes at micro- and nano-scales. Applied Physics Reviews 2015, 2, 011303.

[13] Ponsinet V., Baron A., Pouget E., Okazaki Y., Oda R. and Barois P. Self-assembled nanostructured metamaterials. EPL 2017, 119, 14004.

[14] Baron A., Aradian A., Ponsinet V. and Barois P. Self-assembled optical metamaterials. Optics & Laser Technology, 2016, 82, 94-100.

[15] Malassis L., Massé P., Treguer-Delapierre M. et al. Bottom-up Fabrication and Optical Characterization of Dense Films of Meta-Atoms Made of Core−Shell Plasmonic Nanoparticles. Langmuir 2013, 29, 1551−1561.

[16] Mühlig S., Cunningham A., Scheeler S. et al. Self-Assembled Plasmonic Core-Shell Clusters with an Isotropic Magnetic Dipole Response in the Visible Range. ACS Nano, 2011, 5, 6586-6596.

[17] Sheikholeslami S.N., Alaeian H., Koh A.L. and Dionne J.A. A Metafluid Exhibiting Strong Optical Magnetism. Nano Lett., 2013, 13, 4137-4141.

[18] Ponsinet V., Barois P., Gali S.M. et al., Resonant isotropic optical magnetism of plasmonic nanoclusters in visible light. Phys. Rev. B 2015, 92, 220414(R)

[19] Qian Z., Hastings S.P., Li C. et al. Raspberry-like Metamolecules Exhibiting Strong Magnetic Resonances. ACS Nano, 2015, 9, 1263-1270.

[20] Gomez-Graña S., Le Beulze A., Treguer-Delapierre M. et al. Hierarchical self-assembly of a bulk metamaterial enables isotropic magnetic permeability at optical frequencies. Mater. Horiz. 2016, 3, 596-601.

[21] Li C., Lee S., Qian Z., Woods C., Park S.-J. and Fakhraai Z. Controlling Magnetic Dipole Resonance in Raspberry-like Metamolecules. J. Phys. Chem. C 2018, 122, 6808−6817.

[22] Dezert R., Richetti P. and Baron A. Isotropic Huygens dipoles and multipoles with colloidal particles. Phys. Rev. B 2017, 96, 180201(R).

[23] Pendry J.B., Negative Refraction Makes a Perfect Lens. Phys. Rev. Lett. 2000, 85, 3966-3969.

[24] Mahmoud A.M. and Engheta N. Wave–matter interactions in epsilon-and-mu-near-zero structures. Nature Commun. 2014, DOI: 10.1038/ncomms663.

[25] Evlyukhin A.B., Novikov S.M., Zywietz U. et al. Demonstration of Magnetic Dipole Resonances of Dielectric Nanospheres in the Visible Region. Nano Lett. 2012, 12, 3749−3755.

[26] Zywietz U., Evlyukhin A.B., Reinhardt C. and Chichkov B.N. Laser printing of silicon nanoparticles with resonant optical electric and magnetic responses. Nature Commun. 2014, DOI: 10.1038/ncomms4402.

[27] De Marco M.L., Semlali S., Korgel B.A., Barois P., Drisko G.L. and Aymonier C. Silicon-Based Dielectric Metamaterials: Focus on the Current Synthetic Challenges. Angew. Chem. Int. Ed. 2018, 57, 2 – 23.

[28] Chomette C., Tréguer-Delapierre M., Schade N.B. et al. Colloidal Alchemy:Conversion of Polystyrene Nanoclusters into Gold. ChemNanoMat 2017, 3,160–163.

[29] Mackowski D.W. and Mishchenko M.I. Calculation of the T matrix and the scattering matrix for ensembles of spheres. J. Opt. Soc. Am. A, 1996, 13, 2266-2278. http://www.eng.auburn.edu/users/dmckwski/scatcodes/

[30] Bohren C.F. and Huffman D.R. Absorption and Scattering of Light by Small Particles (Wiley-Interscience, New York, 1983).

[31] Sharma N.L. Nondipole Optical Scattering from Liquids and Nanoparticles. Phys. Rev. Lett. 2007, 98, 217402.

[32] Gomez-Graña S., Treguer-Delapierre M., Duguet E. et al. Isotropic 3D Optical Magnetism in Visible Light in a Self-Assembled Metamaterial. IEEE Xplore 2016, DOI: 10.1109/MetaMaterials.2016.7746432.

[33] Jain P.K., Wenyu Huang and El-Sayed M.A. On the Universal Scaling Behavior of the Distance Decay of Plasmon Coupling in Metal Nanoparticle Pairs: A Plasmon Ruler Equation. Nano Lett. 2007, 7, 2080-2088.

[34] Grahn P., Shevchenko A. and Kaivola M. Electromagnetic multipole theory for optical nanomaterials. New Journal of Physics, 2012, 14(9), 093033.

[35] Amendola V., Bakr O. M. and Stellacci, F. A study of the surface plasmon resonance of silver nanoparticles by the discrete dipole approximation method: effect of shape, size, structure, and assembly. Plasmonics 2010, 5(1), 85-97.